\documentclass[copyright, creativecommons]{eptcs}
\providecommand{\event}{WWV 2014}

\usepackage{breakurl} 
\usepackage{makeidx}
\usepackage{graphicx,latexsym,alltt}
\usepackage{amssymb,amsmath}
\usepackage{algorithm,algorithmic}
\usepackage{subfigure}
\usepackage{color}
\usepackage{url}

\newcommand{\figures}[1]{./figures/#1}
\newcommand{\biblio}[1]{./Biblio/#1}

\definecolor{blank}{rgb}{0.7,0.7,0.7}

\newcounter{contadorDefiniciones}
\newcounter{contadorEjemplos}

\newtheorem{definition}[contadorDefiniciones]{Definition}
\newtheorem{example}[contadorEjemplos]{Example}

\long\def\comment#1{}

\renewcommand{\phi}{\varphi}

\def\defemb#1#2{\expandafter\def\csname #1\endcsname
                              {\relax\ifmmode #2\else\hbox{$#2$}\fi}}
\defemb{cA}{{\cal A}}
\defemb{cB}{{\cal B}}
\defemb{cC}{{\cal C}}
\defemb{cD}{{\cal D}}
\defemb{cE}{{\cal E}}
\defemb{cF}{{\cal F}}
\defemb{cG}{{\cal G}}
\defemb{cH}{{\cal H}}
\defemb{cI}{{\cal I}}
\defemb{cJ}{{\cal J}}
\defemb{cL}{{\cal L}}
\defemb{cM}{{\cal M}}
\defemb{cN}{{\cal N}}
\defemb{cO}{{\cal O}}
\defemb{cP}{{\cal P}}
\defemb{cR}{{\cal R}}
\defemb{cS}{{\cal S}}
\defemb{cT}{{\cal T}}
\defemb{cU}{{\cal U}}
\defemb{cV}{{\cal V}}
\defemb{cW}{{\cal W}}
\defemb{cX}{{\cal X}}
\defemb{cZ}{{\cal Z}}

\makeatletter
\newenvironment{prog}{\vspace{1.0ex}\par
\obeylines\@vobeyspaces\tt}{\vspace{1.0ex}\noindent
}
\makeatother
\newcommand{\startprog}{\begin{prog}}
\newcommand{\stopprog}{\end{prog}\noindent}
\newcommand{\pr}[1]{\mbox{\tt #1}}   

\catcode`\_=\active
\let_=\sb
\catcode`\_=12
\makeatother

\title{Automatic {D}etection of {W}ebpages\\ that {S}hare the {S}ame {W}eb {T}emplate}
\author{Juli\'an Alarte \qquad David Insa \qquad Josep Silva
\institute{Universitat Polit\`ecnica de Val\`encia\\ Valencia, Spain}
\email{\{jalarte,dinsa,jsilva\}@dsic.upv.es}
\and
Salvador Tamarit
\institute{Universidad Polit\'ecnica de Madrid\\
Madrid, Spain}
\email{stamarit@babel.ls.fi.upm.es}
}
\def\titlerunning{Automatic {D}etection of {W}ebpages that {S}hare the {S}ame {W}eb {T}emplate}
\def\authorrunning{J. Alarte, D. Insa, J. Silva \& S. Tamarit}

\begin{document}
\maketitle


\begin{abstract}
Template extraction is the process of isolating the template of a given webpage. It is widely used in several disciplines, including webpages development, content extraction, block detection, and webpages indexing. One of the main goals of template extraction is identifying a set of webpages with the same template without having to load and analyze too many webpages prior to identifying the template.
This work introduces a new technique to automatically discover a reduced set of webpages in a website that implement the template. This set is computed with an hyperlink analysis that computes a very small set with a high level of confidence.
\end{abstract}


\section{Introduction}
Internet is full of web templates (in the following just template). Web developers normally use templates to equip their webpages with a common vocabulary of colors, panels and menus. 
Templates are prepared HTML pages where formatting is already implemented and visual components are ready to insert content. Thus, they speed up the development process and they are also useful for the automatic generation of webpages that only need to be fed with content. 

Templates allow developers to compose their webpages with independent blocks that can be reused. This is good for web development because many tasks can be automated and webpage sections can be maintained separately. In fact, many webpage development environments and code generators offer collections of templates that already include Javascript,  CSS, Flash, etc. 
Templates are also good for users, who can benefit from intuitive and uniform designs with a common look and feel and colored and formatted visual elements that increase the navigability and usability of the webpage.

The importance of templates also affects crawlers and indexers, because they usually judge the relevance of a webpage according to the frequency and distribution of terms and hyperlinks. 
Since templates contain a considerable number of
common terms and hyperlinks that are replicated in a large
number of webpages, relevance may turn out to be inaccurate, leading to incorrect results (see, e.g., \cite{BarR02,VieSPMCF06,YiLL03}). 
Moreover, in general, templates do not contain relevant content, they usually contain one or more pagelets \cite{Cha01,BarR02} (i.e., self-contained logical regions with a well defined topic or functionality) where the main content must be inserted. 
Therefore, detecting templates can allow indexers to identify the main content of the webpage. 

Modern crawlers and indexers do not treat all terms in a webpage in the same way. Webpages are preprocessed to identify the template because template extraction allows them to identify those pagelets that only contain noisy information such as advertisements and banners. This content should not be indexed in the same way as the relevant content. Indexing the non-content part of templates not only affects accuracy, it also affects performance and can lead to a waste of storage space, bandwidth and time.

Template extraction enhance indexers by isolating the main content and assigning higher weights to the really relevant terms.
Once templates have been extracted, they are processed for indexing---they can be analyzed only once for all webpages using the same template---. Moreover, links in templates allow indexers to discover the topology of a website (e.g., through navigational content such as menus), thus identifying the main webpages. They are also essential to compute pageranks.

Gibson et al. \cite{GibPT05} determined that templates represent between 40\% and 50\% of data on the Web and that around 30\% of the visible terms and hyperlinks appear in templates. This justifies the importance of template removal \cite{YiLL03,VieSPMCF06} for web mining and search. 

Given a webpage $w$, conceptually, template extraction is made in two steps:\footnote{In practice, these two steps are not necessarily sequential. In fact, they are often interlaced.}

\begin{enumerate}
\item Detect a set $S$ of webpages that implement the same template than $w$.
\item Analyze all webpages in $S\cup\{w\}$ to identify the template.
\end{enumerate} 

Our technique solves the first step through an analysis of the hyperlinks in the webpages.
We introduce a new idea to automatically find a set of webpages that potentially share a template. 
Roughly, we detect the menu and analyze its links to identify a set of mutually linked webpages. 
One of the main functions of a template is in aiding navigation, thus almost all templates provide a large number of links, shared by all webpages implementing the template. Locating the menu allows us to identify in the topology of the website the main webpages of each category or section. These webpages very likely share the same template. 
This idea is simple but powerful and, contrarily to other approaches, it allows the technique to only analyze a reduced set of webpages to identify the template.

In practice, not all webpages in a website implement the whole template. They often implement a part of the template. As a consequence, our hyperlink analysis has two apparently contradictory objectives: (i) finding webpages as similar as possible to the target webpage, so that it is easy to identify the template, and (ii) finding webpages as different as possible between them, so that we have an heterogeneous sample that implements different parts of the template.   

The rest of the paper has been structured as follows:
In Section~\ref{sec_rel} we discuss the state of the art and compare other approaches. 
Then, in Section~\ref{sec_temex}, we present our technique with examples and explain the algorithms used. 
In Section~\ref{sec_impl} we give some details about the implementation.
Finally, Section~\ref{sec_concl} concludes.

\section{Related Work}\label{sec_rel}

Template extraction techniques are often classified into two groups: page-level and site-level.
In both cases, the objective is the same, detecting the template of a given webpage; but they use different information. While page-level techniques only use the information contained in the target webpage, site-level techniques also use the information contained in other webpages, typically of the same website. 

Site-level techniques usually work in two (not necessarily independent) phases. First, they collect a set of webpages that (hopefully) implement the same template as the target webpage. Then, they extract the template by comparing the target webpage with the collected webpages. 
Our technique automates the first step. Despite there are many site-level template extraction techniques in the literature, there are very few approaches that describe a methodology to detect webpage candidates, and very often, this process is done manually. 

There exist three main different approaches to template extraction, namely, (i) using the textual information of the webpages (i.e., the HTML code), (ii) using the rendered images of the webpages in the browser, and (iii) using the DOM trees of the webpages. 

The first approach is based on the idea that the main content of the webpage has more density of text, with less labels. For instance, the main content can be identified selecting the largest contiguous text area with
the least amount of HTML tags \cite{FerZBB08}. This has been measured directly on the HTML code by counting the number of characters inside text, and characters inside labels. This measure produce a ratio called CETR \cite{WenHH10} used to discriminate the main content. Other approaches exploit densitometric features based on the observation that some specific terms are more common in templates \cite{KohN08,Koh09}. 

The second approach assumes that the main content of a webpage use to be in the central part and (at least partially) visible without scrolling \cite{BurR09}. This approach has been less studied because rendering webpages for classification is a computational expensive operation \cite{KohFN10}. 

The third approach analyzes the attributes and relative positions of DOM nodes. While some works try to identify pagelets analyzing the DOM tree with heuristics \cite{BarR02}, others try to find common subtrees in the DOM trees of a collection of webpages in the website \cite{YiLL03,VieSPMCF06}.  

With independence of the approach followed, the most extended way of selecting the webpage candidates is manually. 
For instance, the content extractor algorithm and its improved version, the fast content extractor algorithm \cite{NguNPB09}, take as input a set of webpages that are given by the programmer. The same happens in the methodology for template extraction proposed in \cite{KadD12}.

Even though \cite{YiLL03} uses a method for template extraction, its main goal is to remove redundant parts of a website. 
For this, they use the Site Style Tree (SST), a data structure that is constructed by analyzing a set of DOM trees and recording every node found, so that repeated nodes are identified by using counters in the SST nodes. Hence, an SST summarizes a set of DOM trees. After the SST is built, they have information about the repetition of nodes. The most repeated nodes are more likely to belong to a noisy part that is removed from the webpages. 
Their technique inputs a collection of webpages to construct the SST. They do not have a methodology to select the webpages, and they do not propose a number of webpages needed. In their experiments, they randomly sample 500 webpages, and the time taken to build a SST is always below 20 seconds. 

The approach in \cite{VieSPMCF06} is based on discovering optimal mappings between DOM trees. 
This mapping relates nodes that appear in more than one webpage, and thus they are considered redundant.
Their technique uses the RTDM-TD algorithm to compute a special kind of mapping called \emph{restricted top-down mapping} \cite{ReiGSL04}. 
In order to select the webpages of the website that should be mapped to identify the template, they pick random webpages until a threshold is reached. In their experiments, they approximated this threshold as a
few dozen of webpages. They need 25 webpages to reach a 0.95 F1 measure.
Contrarily, in our technique, we do not select the webpages randomly, we use a method to identify the webpages by analyzing their hyperlinks. 
We only need to explore a few webpages to identify the candidates that implement the template.
Moreover, contrarily to us, they assume that all webpages in the website share the same template, and this is a strong limitation for many websites. 

In \cite{RowlandsTW09}, authors exploit the idea that those pages stored in the same directory contain the same template. In particular, they use as webpage candidates those webpages stored in the same directory as the target webpage. Somehow, we also exploit this idea, but we do not restrict ourselves to one directory. We define an order of relevance using the tree of directories according to a definition of distance between directories. 

\section{Identifying webpages that implement the same template} \label{sec_temex}

\begin{figure}[t]
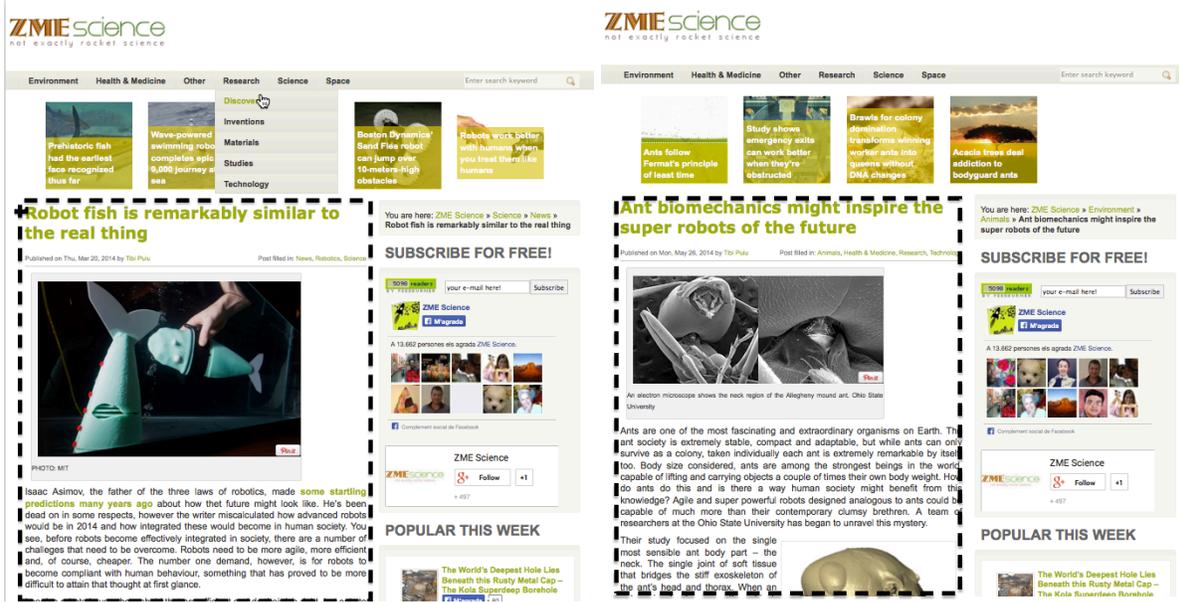

	\centering
		\includegraphics[width=0.49\textwidth]{\figures{pagelets1n_wwv14.pdf}}
		\includegraphics[width=0.49\textwidth]{\figures{pagelets2n_wwv14.pdf}}
	\caption{Webpages of ZMEscience sharing a template}
	\label{fig-pagelets}
\end{figure}

Templates are often composed of a set of pagelets. 
Two of the most important pagelets in a webpage are the menu and the main content. 
For instance, 
in Figure~\ref{fig-pagelets} we see two webpages that belong to the ZMEscience website. At the top of the webpages we see the main menu containing links to all ZMEscience principal topics. In the left webpage we can also see an example of a submenu showing the subsections of topic ``Research". The left webpage belongs to ``Robotics" subsection of topic ``Science", while the right webpage belongs to ``Animals" section of topic ``Environment". Both share the same menu, their respective submenus, and general structure. 
In both webpages the main content, i.e., the news, is inside the pagelet in the dashed square. 
In addition to the main content, there is a common pagelet called ``Popular this week" with the most relevant news, and another one for subscription and social networks. Additionally, a set of related news (different for each webpage) is shown between the menu and the main content.

Our technique inputs a webpage (called key page) and it outputs a set of webpages that implement (a part of) the same template. To discover these webpages, it identifies a complete subdigraph in the website topology.

\subsection{Complete subdigraphs}

Given a website topology, a complete subdigraph (CS) represents a collection of webpages that are pairwise mutually linked. A n-complete subdigraph (n-CS) is formed by n nodes.  
Our interest in complete subdigraphs comes from the observation that the webpages linked by the items in a menu usually form a CS. This is a new way of identifying the webpages that contain the menu. At the same time, these webpages are the roots of the sections linked by the menu. 
The following example illustrates why menus provide very useful information about the interconnection of webpages in a given website. 

\begin{example}
Consider the ZMEscience website. Two of its webpages are shown in Figure~\ref{fig-pagelets}. 
In this website all webpages share the same template, and this template has a main menu that is present in all webpages, and a submenu for each item in the main menu. 
The site map of the ZMEscience website may be represented with the topology shown in Figure~\ref{fig-topology}. 

\begin{figure}[h]
	\centering
		\includegraphics[width=0.49\textwidth]{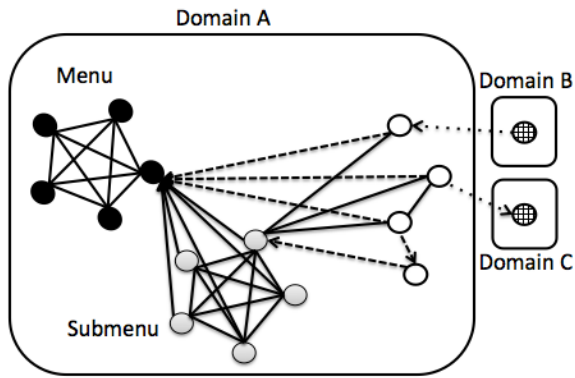}
	\caption{ZMEscience Website topology}
	\label{fig-topology}
\end{figure}

In this figure, each node represents a webpage and each edge represents a link between two webpages (we only draw some of the edges for clarity). Solid edges are bidirectional, and dashed and dotted edges are directed. 
Black nodes are the webpages pointed by the main menu.
Because the main menu is present in all webpages, then all nodes are connected to all black nodes.   
Therefore all black nodes together form a complete graph (i.e., there is an edge between each pair of nodes). 
Grey nodes are the webpages pointed by a submenu, thus, all grey nodes together also form a complete graph. White nodes are webpages inside one of the sections of the submenu, therefore, all of them have a link to all black and all grey nodes.
\end{example}

Of course, not all webpages in a website implement the same template, and some of them only implement a subset of a template. For this reason, one of the main problems of template extraction is deciding what webpages should be analyzed. Minimizing the number of webpages analyzed is essential to reduce the web crawlers work. 
In our technique we introduce a new idea to select the webpages that must be analyzed: we identify a menu in the key page and we analyze the webpages pointed out by this menu. Observe that we only need to investigate the webpages linked by the key page, because they will for sure contain a CS that represents the menu.

In order to ensure high precision, we search for a CS that contains enough webpages that implement the template. It is important to remark that a webpage can contain several menus and submenus; and not all of them produce equally good CSs. For instance, consider again the topology shown in Figure~\ref{fig-topology}. If we assume that the key page is one of the white nodes, then, a CS formed with the grey nodes (the submenu) will be probably better than a CS formed with the black nodes (the main menu). This happens because the white nodes belong to one of the items in the submenu, and thus, they (probably) are more related semantically, and they (probably) share more syntax components. Note that at least the submenu is a common substructure shared by all grey and white nodes. 

\subsection{Hyperlink analysis}

By analyzing the links in the key page, it is possible to select those links that most likely produce the best CS. 
This is essential to avoid analyzing all links and thus significantly increasing the performance. 
Our strategy to identify the links that should be analyzed is based on the structure of the website. We obtain information about the structure of the website from the URLs of the links. 

\begin{example}
\label{ex_URLs}
Consider a key page $P$ whose URL is:\\ 
\url{www.upv.es/research/maths/index.html}\\ 
Consider that $P$ contains four links with the following URLs:
\begin{itemize}
\item URL 1 =  \url{www.tesco.es/}
\item URL 2 =  \url{www.upv.es/research/maths/pi.html}
\item URL 3 =  \url{www.upv.es/sport/}
\item URL 4 =  \url{www.upv.es/research/maths/news/computers.html}
\end{itemize}
\begin{description}
\item [URL 1] points to a webpage in another domain. Therefore, the template of this webpage is probably completely different from the template of the key page.  
\item [URL 2] points to a webpage in the same directory as the key page. Hence, very likely, they both belong to the same section in the hierarchy of the website, and thus their structure is probably similar.
\item [URL 3] points to a webpage (index.html) inside a directory that is two levels above the current directory. It probably points to another section (e.g., to another section in the main menu called \emph{sport}). Therefore, the structures of the key page and the webpage pointed by URL 3 are possibly different. Probably, they will only share a small part of their templates.
\item [URL 4] points to a webpage located inside a subdirectory of the reference directory. Probably, this webpage is semantically related to the key page, and it contains specialized information (it possibly extends the template with additional information).   
\end{description}
\end{example}

Therefore, by analyzing the links in the key page, we can establish an order of relevance. 
To formally define a partial order we need first to provide a notion of link and distance between links. 

\begin{definition}[path, hyperlink]
A \emph{path} is a sequence of words joined by juxtaposition, $s=w_1w_2w_3...w_n$,  
the length of the sequence is represented with $|s|$ and it denotes the number of words in the sequence.
A \emph{hyperlink} (or just \emph{link}) is a path where each word finishes with slash: $h = (dir/)^+$.
\end{definition}


Note that this definition of hyperlink deliberately ignores the name of the resource pointed by the URL; it only focusses on the structure (directories or domains). 
It is general enough as to include URLs such as  $www.upv.es/$, $research/$ and  $research/maths/$.
We use function $head$ to select the first word (i.e., directory) of a hyperlink: $head(dir1/dir2/dir3/)=dir1$.
We now provide a notion of distance between two URLs. 

\begin{definition}[hyperlink distance] \label{def-hyperlinkdistance}
Given two hyperlinks $h, h'$, the distance from $h$ to $h'$ is defined as:
	\begin{equation}
	\label{eq_distance}
		hDistance (h, h') = \left\{
			\begin{array}{ll}
				0      & \mathrm{if\ } h = h'\\
				+|h_1| & \mathrm{if\ } h' = h h_1\\
				-|h_1| & \mathrm{if\ } h = h' h_1\\		 
				-|h_1| & \mathrm{if\ } h = h_0 h_1 \mathrm{\ and\ } h' = h_0 h_2 \mathrm{\ and\ } head(h_1)\neq head(h_2)\\
				-|h| & \mathrm{if\ } head(h)\neq head(h')

			\end{array}
		\right.
	\end{equation}
\end{definition}

\begin{figure}[h]
	\centering
		\includegraphics[width=0.3\textwidth]{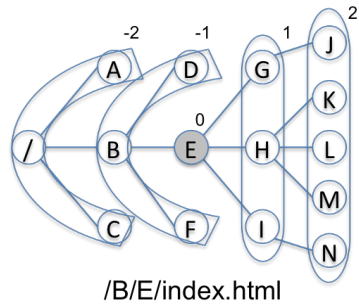}
	\caption{Hyperlink distance}
	\label{fig-hyperlink_distance}
\end{figure}

Note that the distance is defined from the first link to the second link. 
This can be observed in Figure~\ref{fig-hyperlink_distance}, which represents a tree of directories that contain webpages. 
There, we can see the distance of all webpages to the webpage in the gray directory.
Some particular examples follow:\\

\indent hDistance(research/maths/, research/maths/) = 0\\
\indent hDistance(research/maths/, research/maths/geometry/) = +1\\
\indent hDistance(research/maths/, research/) = -1\\
\indent hDistance(research/maths/, research/physics/dynamics) = -1\\
\indent hDistance(research/maths/, www.upv.es/research/) = -2\\


Intuitively, given two links $h1$ and $h2$, a distance of 0 means that both links point to the same directory. 
A positive distance from $h1$ to $h2$, means that $h2$ points to a subdirectory of the directory pointed by $h1$. 
A negative distance from $h1$ to $h2$, means that $h2$ points to a directory outside the directory pointed by $h1$.     
We use the url of the key page as a reference link to compute distances. And we compare the distances of the links of the key page.
Those links with distance 0 are preferred. Then, those with a positive distance. And finally, those with a negative distance. 

In case of a draw\footnote{Note that, according to Definition~\ref{def-hyperlinkdistance}, the hyperlink distance defines a partial order, and thus, two different links can have the same distance to a third link.}, we use another information to determine what link is better. 
Concretely, we analyze their position in the DOM tree. Often, pagelets agglutinate semantically related information. Thus, different pagelets contain different information. Therefore, two links that belong to different pagelets usually point to webpages whose content is semantically different. 
This is very useful, because we are interested in locating webpages that share the same template, and that contain as more differences as possible so that we can precisely identify the template. 

\begin{example}
Consider a set of links in the menu of a shopping webpage. 
The links can point to similar webpages where each webpage contains information about one particular product. 
Often, these webpages share the same template that is filled with similar information about the products. 
Some of the information shared by the products can be confused as part of the template because it appears repeated in many webpages. 
Hence, template extraction algorithms must avoid comparing only these webpages because they do not provide sufficient information to isolate the template.
\end{example}

As a consequence, in case of a draw, we prefer those links that 
are as separated as possible from the other already selected links in the DOM tree. 
In this way, we give preference to links with (probably) different semantic information. 
In summary, observe that we obtain webpages that share the same template (using the hyperlink distance) but being as different as possible (using their position in the DOM tree).

\begin{figure}[h!]
	\centering
		\includegraphics[width=0.7\textwidth]{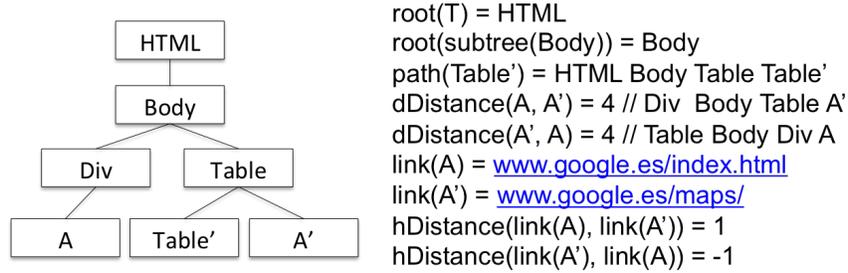}
	\caption{A DOM tree T}
	\label{fig-mapping}
\end{figure}

In the following, webpages are represented with a DOM tree $T = (N, E)$ 
(see Figure~\ref{fig-mapping}). 
$root(T)$ denotes the root node of $T$. 
Given a node $n \in N$, 
$link(n)$ denotes the hyperlink of $n$ when $n$ is a node that represents a hyperlink (HTML label {\tt <a>}).

To compare the position in the DOM tree of two links we measure the length of their DOM paths. 
The DOM path of a node $n$, $path(n)$ is the path from the root to that node.



\begin{definition}[DOM distance]
Given a DOM tree $T = (N, E)$ and two nodes $n, n' \in N$, the \emph{DOM distance} from $n$ to $n'$ is defined as:
	\begin{equation}
	\label{eq_nodesDistance}
		dDistance(n, n') = \left\{
			\begin{array}{ll}
				0      & \mathrm{if\ } path(n) = path(n')\\
				j + k & \mathrm{if\ } path(n) = n_0...n_i m_1...m_j~
					\mathrm{ and\ } path(n') = n_0...n_i m'_1...m'_k~
					\mathrm{ and\ } m_1 \neq m'_1\\
			\end{array}
		\right.
	\end{equation}
\end{definition}

%
%
%

Note that two links have zero DOM distance if and only if they are exactly the same link. 
Contrarily, two different links (even if they have the same URL, and thus the same hyperlink distance) necessarily have a positive DOM distance.

We can now define an order for the links in a webpage. It allows us to decide what links should be explored to extract the template. This order combines the link relevance order $\leq_{link}^{h}$, which uses the link distance, and the DOM relevance order $\leq_{\mathit{DOM}}^{N}$, which uses the DOM distance. Their formal definition follows.

\begin{definition}[link relevance]
Given any set of hyperlink nodes $N$ of a DOM tree and a reference hyperlink $h$, $N$ is equipped with the preorder $\leq_{link}^{h}$ called \emph{link relevance} and defined as follows.  
For any $n_1, n_2 \in N$ we have:
\begin{center}
\underline{Link Relevance:}\\
\medskip
$n_1 =_{link}^{h} n_2  $ ~~~iff~~~ $ hd_1 = hd_2$\\
\bigskip

$n_1 <_{link}^{h} n_2  $ ~~~iff~~~ $ \left\{
	\begin{array}{ll}
		0 \leq hd_1 < hd_2 ~\vee\\
		hd_2 < hd_1 \leq 0 ~\vee\\
		hd_2 < 0 \leq hd_1
	\end{array}
\right.
$
\end{center}

where

\smallskip
\indent~~~~~~~~~~~~~~~~~~~~~ $hd_1 = hDistance(h, link(n_1))$ \hspace{4em} $hd_2 = hDistance(h, link(n_2))$

\end{definition}

\begin{definition}[DOM relevance]
Given any set of hyperlink nodes $N$ of a DOM tree $T$ and a reference set of hyperlink nodes $N'$ in $T$, $N$ is equipped with the preorder $\leq_{\mathit{DOM}}^{N'}$ called \emph{DOM relevance} and defined as follows.  
For any $n_1, n_2 \in N$ we have:
\begin{center}
\underline{DOM Relevance:}\\
\medskip
%
$n_1 =_{\mathit{DOM}}^{N'} n_2$ ~~~iff~~~ $ \left\{
	\begin{array}{ll}
		N'=\emptyset ~\vee\\
		dn_1' = dn_2'
	\end{array}
\right.
$\\
\bigskip

$n_1 <_{\mathit{DOM}}^{N'} n_2$ ~~~iff~~~ $dn_1' > dn_2'$  
\end{center}

where

\smallskip
\indent$~~~~~~~~~~~~~~~~~~~~~~~~~~~~~$ $n_1' = \min\limits_{n \in N'} dDistance(n, n_1)$ \hspace{4em} ~$dn_1' = dDistance(n_1', n_1)$\\
\indent$~~~~~~~~~~~~~~~~~~~~~~~~~~~~~$ $n_2' = \min\limits_{n \in N'} dDistance(n, n_2)$ \hspace{4em} $dn_2' = dDistance(n_2', n_2)$

\end{definition}

\subsection{Finding webpage candidates in a website}

We use a combination of link relevance and DOM relevance to select the links that should be explored first to find a CS in the website topology. For this, we use Algorithm~\ref{alg_sortLinks}.

\begin{algorithm}[h!]
\caption{Sort links}
\label{alg_sortLinks}

\begin{algorithmic}
{\scriptsize
\smallskip
\STATE \textbf{Input:} A set of hyperlink nodes $links$ and a reference hyperlink $h$.
\STATE \textbf{Output:} A sorted list of $links$ with respect to the preorders $\leq_{\mathit{link}}^{h}$ and $\leq_{\mathit{DOM}}^{N}$.\\
\medskip

\STATE $\textbf{begin}$
\STATE $~~\mathit{sortedLinks} = [ ]$;
\STATE ~~\textbf{while} ($\mathit{links}$ $\neq$ $\emptyset$)
\STATE $~~~~~~\mathit{links'} = \{ \mathit{l} \in \mathit{links} \mid \nexists \mathit{l'} \in \mathit{links} \land \mathit{l'} <_{\mathit{link}}^{h} \mathit{l} \}$;
\STATE $~~~~~~\mathit{links} = \mathit{links} \backslash \mathit{links'}$;
\STATE $~~~~~~\mathit{sortedLinks'} = [ ]$;
\STATE ~~~~~~\textbf{while} ($\mathit{links'}$ $\neq$ $\emptyset$)
\STATE $~~~~~~~~~~\mathit{link} = l \in \mathit{links'} \mid \nexists \mathit{l'} \in \mathit{links'} \land l' <_{\mathit{DOM}}^{\mathit{sortedLinks'}} l$;
\STATE $~~~~~~~~~~\mathit{links'} = \mathit{links'} \backslash \{ \mathit{link} \}$;
\STATE $~~~~~~~~~~\mathit{sortedLinks'} = \mathit{sortedLinks'}$ $\mathit{++}$ $[ \mathit{link} ]$;
\STATE $~~~~~~\mathit{sortedLinks} = \mathit{sortedLinks}$ $\mathit{++}$ $\mathit{sortedLinks'}$;
\STATE ~~\textbf{return} $\mathit{sortedLinks}$;
\STATE $\textbf{end}$

}
\end{algorithmic}
\end{algorithm}

Algorithm~\ref{alg_sortLinks} sorts the links in a webpage combining link relevance and DOM relevance. First, it takes each set of hyperlink nodes in the order provided by the link relevance. Then, it sorts each of these sets using the order provided by the DOM relevance. Finally, the concatenation of each sorted set is the order that we use to explore the links of a webpage.

Now we are in a position to describe our algorithm that identifies a CS in a website. 
This algorithm is Algorithm~\ref{AlgoMCS}.

\begin{algorithm}[h!]
\caption{Extract a n-CS from a website}
\label{AlgoMCS}

\begin{algorithmic}
{\scriptsize
\smallskip
\STATE \textbf{Input:} An $initialLink$ that points to a webpage and the expected size $n$ of the CS.
\STATE \textbf{Output:} A set of links to webpages that together form a n-CS.\\ 
$~~~~~~~~~~~~~$If a n-CS cannot be formed, then they form the biggest m-CS with m $<$ n.
\medskip

\STATE $\textbf{begin}$
\STATE $~~\mathit{keyPage} = $ $\mathit{loadWebPage}$$(\mathit{initialLink})$;
\STATE $~~\mathit{reachableLinks} =  \mathit{getLinks} (\mathit{keyPage})$; 
\STATE $~~\mathit{processedLinks} = \emptyset$;
\STATE $~~\mathit{connections} = \emptyset$;
\STATE $~~\mathit{bestCS} = \emptyset$;
\STATE $~~\mathit{sortedLinks} =  \mathit{sortLinks} (\mathit{reachableLinks},\mathit{initialLink})$;
\STATE ~~\textbf{foreach} ($link$ \textbf{in} $\mathit{sortedLinks}$)
\STATE $~~~~~~\mathit{webPage} =  \mathit{loadWebPage} (link)$;
\STATE $~~~~~~\mathit{existingLinks} =  \mathit{getLinks} (webPage) \cap \mathit{reachableLinks}$;
\STATE $~~~~~~\mathit{processedLinks} = \mathit{processedLinks} \cup \{ link \}$;
\STATE $~~~~~~\mathit{connections} = \mathit{connections} \cup \{(link \rightarrow \mathit{existingLink}) \mid \mathit{existingLink} \in \mathit{existingLinks}\}$; 
\STATE $~~~~~~CS = \{ ls \in \cP(\mathit{processedLinks}) \mid link \in ls \land \forall l, l' \in ls~.~(l \rightarrow l'),(l' \rightarrow l) \in \mathit{connections} \}$;
\STATE $~~~~~~\mathit{maximalCS} = cs \in CS$ such that $\forall cs' \in CS~.~|cs| \geq |cs'|$;
\STATE ~~~~~~\textbf{if} $|\mathit{maximalCS}| = n$ \textbf{then} \textbf{return} $\mathit{maximalCS}$;
\STATE ~~~~~~\textbf{if} $|\mathit{maximalCS}| > |\mathit{bestCS|}$ \textbf{then} $\mathit{bestCS} = \mathit{maximalCS}$;
\STATE ~~\textbf{return} $\mathit{bestCS}$;
\STATE $\textbf{end}$


}
\end{algorithmic}
\end{algorithm}

The algorithm uses two trivial functions:
$\mathit{loadWebPage}(link)$, which loads and returns the webpage pointed by the input link, and 
$\mathit{getLinks}(webpage)$, which returns the collection of (non-repeated) links\footnote{In our implementation, this function removes those links that point to other domains because they are very unlikely to contain the same template. Here, we do not impose this restriction to keep the algorithm general.} in the input webpage (ignoring self-links). 
Function $\mathit{sortLinks}$ corresponds to Algorithm~\ref{alg_sortLinks}.
Observe that the main loop iteratively explores the links of the webpage pointed by the $initialLink$ (i.e., the key page) until it founds a n-CS. Note also that it only loads those webpages needed to find the n-CS, and it stops when the n-CS has been found. 
We want to highlight the mathematical expression
\medskip
\begin{center}
\small
$CS = \{ ls \in \cP(processedLinks) \mid link \in ls \land \forall l, l' \in ls~.~(l \rightarrow l'),(l' \rightarrow l) \in connections \}$
\end{center}
\medskip
\noindent where $\cP(X)$ returns all possible partitions of set $X$.

This line is used to find the set of all CS that can be constructed with the current $link$.
The current link must be part of the CS ($link \in ls$) to ensure that we make progress (not repeating the same search of the previous iteration). 
Moreover, because the CS is constructed incrementally, the statement
\medskip
\begin{center}
\small
\textbf{if} $|\mathit{maximalCS}| = n$ \textbf{then} \textbf{return} $\mathit{maximalCS}$
\end{center}
\medskip
\noindent ensures that whenever a n-CS can be formed, it is returned.    

\section{Implementation} \label{sec_impl}

The technique presented in this paper including all algorithms described has been implemented as a Firefox's toolbar. 
We selected Firefox because it is one of the most powerful and widely used browsers, and it is free and open source.
Firefox toolbars are implemented using XUL, an XML based language used to implement the interface; and Javascript, which implements the behavior and actions of the toolbar. 
In total, it contains 2577 LOC.

With this tool, the user can browse the Internet as usual. 
Then, when she wants to extract the template of a webpage, she only needs to press a button and the tool automatically 
loads the appropriate linked webpages to form a CS. The links to all webpages are then displayed in the browser.

\begin{example}

Consider the key page in Figure~\ref{fig-nCS} left. We can introduce in the top bar the number of webpages that implement the same template than the key page, or we can left the default value 3, which produces the best balance between precision and performance according to our experiments. In this case, 4 webpages were required, thus, if we press the button in the top bar, then the webpage at the right is automatically generated. These links point to webpages that together form a 4-CS. All of them are very likely to implement the same template than the key page.

\begin{figure}[t]
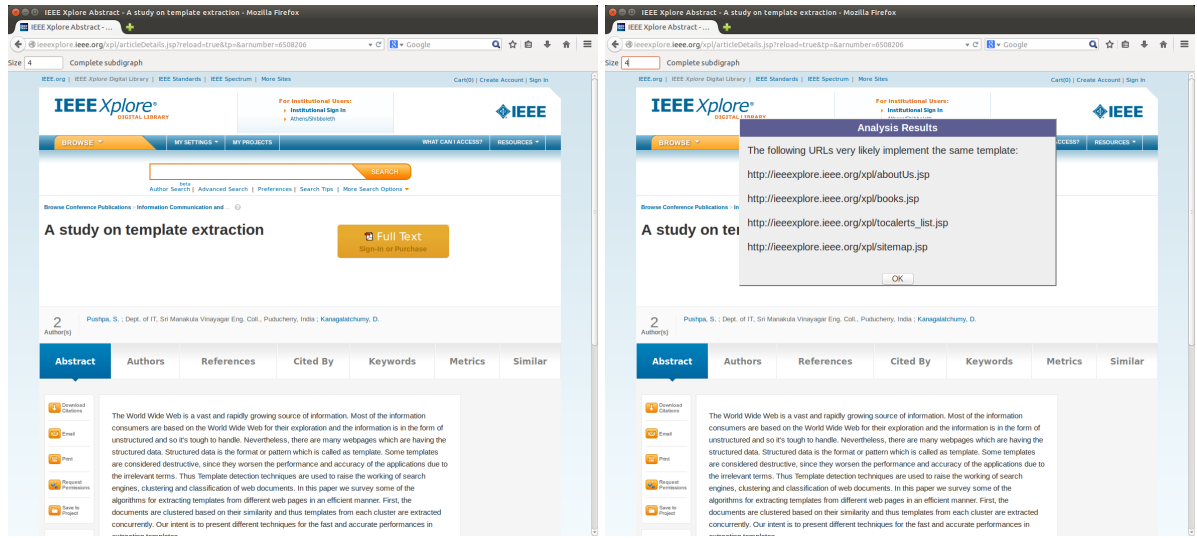

	\centering
		\includegraphics[width=0.49\textwidth]{\figures{keypage.png}}
		\includegraphics[width=0.49\textwidth]{\figures{ncs.png}}
	\caption{Keypage (left) and a set of webpages (right) automatically identified with the tool.}
	\label{fig-nCS}
\end{figure}
\end{example}

\bigskip
\noindent{\bf Empirical evaluation} 
\bigskip

In our theoretical formalization we presented our technique in an abstract way.
Some definitions such as hyperlink distance, DOM distance, and relevance, reveal features of a website that must be considered when extracting templates. Other features, however, have been left as parameters of our algorithms. 
For instance, Algorithm~\ref{AlgoMCS} can compute a CS of any specified size. 
In this section we discuss how to determine the size of the CS based on empirical evaluation.    

One important design decision is related to the domain boundaries of the websites analyzed.
It is possible that several webpages of different domains are mutually linked. 
Sometimes this is even usual between the main webpages of different companies in an alliance. 
They all point to the others, e.g., with a set of logos. 
Nevertheless, the templates of the companies are often different.
In fact, in our experiments, we did not find a shared template between different domains. 
Therefore, for efficiency reasons, external domains are omitted when computing the CS. 
In our implementation, we restrict our search to webpages in the same domain as the key page.

\bigskip
\noindent{\bf Determining the size of the complete subdigraph}
\bigskip

Algorithm~\ref{AlgoMCS} computes a n-CS in a website. 
As previously explained, there are several combinations of webpages that form a CS. 
One could think that the bigger the CS is, the better; so we could even think in calculating the maximal CS. 
Nevertheless, this is not a good idea. Firstly, because computing the maximal CS has an exponential cost. 
And secondly, because experiments reveal that increasing the size of the CS does not necessarily imply a better precision or recall.  

In order to prove this, we performed several experiments with a template extractor that implemented our technique to identify the set of webpages used to identify the template. Then, we repeated all the experiments with different sizes (1,2,3,4,5,6,7 and 8) for the CS in order to determine the best value.
We have made public the details of the experiments including the suit of benchmarks used at:
\begin{center}
\url{http://www.dsic.upv.es/~jsilva/retrieval/templates/experiments.html}. 
\end{center}
The results are presented in Table~\ref{tab:size}.
\begin{table*}[h!]
 {\scriptsize

 \centering
\hspace{4.3cm}
\begin{tabular}{|l|r|r|r|r|r|} \hline
 \texttt{Size}~ & ~\texttt{Recall}~ & ~\texttt{Precision}~ & \texttt{F1~~~}~ & ~\texttt{Loads} \\
 \hline\hline

 \pr{~~1}          & 88,56 \%~  & 94,89 \%~  & ~88,69 \%~  & 2~  \\
 \pr{~~2}          & ~96,34 \%~  & 90,32 \%~  & 91,93 \%~  & 5,6~  \\
 \pr{~~3}          & 95,44 \%~  & 96,35 \%~  & 95,61 \%~  & 10,13~  \\
 \pr{~~4}          & 94,61 \%~  & 96,88 \%~  & 95,27 \%~  & 16,52~  \\
 \pr{~~5}          & 94,69 \%~  & 96,96 \%~  & 95,40 \%~  & 18,68~  \\
 \pr{~~6}          & 95,21 \%~  & 96,82 \%~  & 95,69 \%~  & 23,68~  \\
 \pr{~~7}          & 95,46 \%~  & 96,31 \%~  & 95,57 \%~  & 30~  \\
 \pr{~~8}          & 95,14 \%~  & 96,57 \%~  & 95,54 \%~  & 32,08~  \\

 \hline
\end{tabular}
}
\medskip
\caption{Determining the size of the complete subdigraph}
\label{tab:size}
\end{table*}

This table summarizes several experiments.
Each row in the table is the average of 40 template extractions from 40 different webpages. Each row is the result of repeating the experiments with a different value for n in the n-CS searched by Algorithm~\ref{AlgoMCS}. 
In particular, each column has the following meaning:
\begin{description}
\item [\texttt{Size:}] represents the size of the CS that the algorithm tried to find in the websites. 
In the case that there did not exist a CS of the searched size, then the algorithm used the biggest CS with a size under the specified size (see Algorithm~\ref{AlgoMCS}).
\item [\texttt{Recall:}] shows the number of correctly retrieved nodes divided by the number of nodes in the gold standard.
\item [\texttt{Precision:}] shows the number of correctly retrieved nodes divided by the number of retrieved nodes.
\item [\texttt{F1:}] shows the F1 metric that is computed as $(2*P*R)/(P+R)$ being $P$ the precision and $R$ the recall.
\item [\texttt{Loads:}] represents the average number of webpages loaded to construct the n-CS. 
\end{description}
 
Observe that F1 is stabilized in 95\% with a CS of size 3. Increasing the size of the CS does not significantly increase F1, but it increases the number of pages loaded to construct the (bigger) CS. 
Therefore, we determined that a subdigraph of size 3 is the best option because it keeps almost the best F1 value and it is quite efficient, the algorithm has to load significantly less webpages than in a size bigger than 3.
As a direct consequence, our implementation, by default, stops when a subdigraph of size 3 has been found (but it can be configured to search for a subdigraph of any size (4,5,6, etc.)). 

Our implementation and all the experimentation is public. 
All the information related to the experiments, the source code of
the benchmarks, the plugin, the source code of the tool
and other material can be found at\\
\vspace{-1em}
\begin{center}
\url{http://www.dsic.upv.es/~jsilva/retrieval/templates/}
\end{center}


\section{Conclusions}
\label{sec_concl}

Templates are useful for website developers, for crawlers and for final users. 
This work presents a new technique for template extraction. 
Given a webpage, the technique automatically detects a set of (linked) webpages that very likely implement the same template. This is done by analyzing the links in the menus of the website. 
To the best of our knowledge, the idea of using the menus to locate the template is new, and it allows us to find a set of webpages from which we can extract the template with a reduced amount of pages loaded. This is especially interesting for performance, because loading webpages to be analyzed is expensive, and this part of the process is minimized in our technique. Our implementation and experiments have shown the usefulness of the technique.

For future work, we plan to investigate a strategy to further reduce the amount of webpages loaded with our technique. 
The idea is to directly identify the menu in the key page by measuring the density of links in its DOM tree. 
The menu has probably one of the higher densities of links in a webpage. 
Therefore, our technique could benefit from measuring the links--DOM nodes ratio to directly find the menu in the key page, and thus, a complete subdigraph in the website topology.


\section{Acknowledgements}
\label{sec_acks}
This work has been partially supported by the Spanish \emph{Ministerio de
Econom\'{\i}a y Competitividad (Secretar\'{\i}a de Estado de
 Investigaci\'on, Desarrollo e Innovaci\'on)}
under grant TIN2013-44742-C4-1-R and by the
\emph{Generalitat Valenciana} under grant PROMETEO/2011/052.
David Insa was partially supported
by the Spanish {Ministerio de Eduaci\'on} under FPU grant AP2010-4415.
Salvador Tamarit was partially supported by research project POLCA, Programming
Large Scale Heterogeneous Infrastructures (610686), funded by the
European Union, STREP FP7.

\bibliography{\biblio{biblio}}

\begin{thebibliography}{10}
\providecommand{\bibitemdeclare}[2]{}
\providecommand{\surnamestart}{}
\providecommand{\surnameend}{}
\providecommand{\urlprefix}{Available at }
\providecommand{\url}[1]{\texttt{#1}}
\providecommand{\href}[2]{\texttt{#2}}
\providecommand{\urlalt}[2]{\href{#1}{#2}}
\providecommand{\doi}[1]{doi:\urlalt{http://dx.doi.org/#1}{#1}}
\providecommand{\bibinfo}[2]{#2}

\bibitemdeclare{inproceedings}{BarR02}
\bibitem{BarR02}
\bibinfo{author}{Ziv \surnamestart Bar-Yossef\surnameend} \&
  \bibinfo{author}{Sridhar \surnamestart Rajagopalan\surnameend}
  (\bibinfo{year}{2002}): \emph{\bibinfo{title}{{T}emplate detection via data
  mining and its applications}}.
\newblock In: {\sl \bibinfo{booktitle}{{P}roceedings of the 11th
  {I}nternational {C}onference on {W}orld {W}ide {W}eb ({WWW}'02)}},
  \bibinfo{publisher}{{ACM}}, \bibinfo{address}{{N}ew {Y}ork, {NY}, {USA}}, pp.
  \bibinfo{pages}{580--591}, \doi{10.1145/511446.511522}.

\bibitemdeclare{inproceedings}{BurR09}
\bibitem{BurR09}
\bibinfo{author}{Radek \surnamestart Burget\surnameend} \&
  \bibinfo{author}{Ivana \surnamestart Rudolfova\surnameend}
  (\bibinfo{year}{2009}): \emph{\bibinfo{title}{{W}eb {P}age {E}lement
  {C}lassification {B}ased on {V}isual {F}eatures}}.
\newblock In: {\sl \bibinfo{booktitle}{{P}roceedings of the 1st {A}sian
  {C}onference on {I}ntelligent {I}nformation and {D}atabase {S}ystems
  ({ACIIDS}'09)}}, \bibinfo{publisher}{{IEEE} {C}omputer {S}ociety},
  \bibinfo{address}{{W}ashington, {DC}, {USA}}, pp. \bibinfo{pages}{67--72},
  \doi{10.1109/ACIIDS.2009.71}.

\bibitemdeclare{inproceedings}{Cha01}
\bibitem{Cha01}
\bibinfo{author}{Soumen \surnamestart Chakrabarti\surnameend}
  (\bibinfo{year}{2001}): \emph{\bibinfo{title}{{I}ntegrating the {D}ocument
  {O}bject {M}odel with hyperlinks for enhanced topic distillation and
  information extraction}}.
\newblock In: {\sl \bibinfo{booktitle}{{P}roceedings of the 10th
  {I}nternational {C}onference on {W}orld {W}ide {W}eb ({WWW}'01)}},
  \bibinfo{publisher}{{ACM}}, \bibinfo{address}{{N}ew {Y}ork, {NY}, {USA}}, pp.
  \bibinfo{pages}{211--220}, \doi{10.1145/371920.372054}.

\bibitemdeclare{inproceedings}{FerZBB08}
\bibitem{FerZBB08}
\bibinfo{author}{Adriano \surnamestart Ferraresi\surnameend},
  \bibinfo{author}{Eros \surnamestart Zanchetta\surnameend},
  \bibinfo{author}{Marco \surnamestart Baroni\surnameend} \&
  \bibinfo{author}{Silvia \surnamestart Bernardini\surnameend}
  (\bibinfo{year}{2008}): \emph{\bibinfo{title}{{I}ntroducing and evaluating
  {ukWaC}, a very large web-derived corpus of english}}.
\newblock In: {\sl \bibinfo{booktitle}{{P}roceedings of the 4th {W}eb as
  {C}orpus {W}orkshop ({WAC}-4)}}, pp. \bibinfo{pages}{47--54}.

\bibitemdeclare{inproceedings}{GibPT05}
\bibitem{GibPT05}
\bibinfo{author}{David \surnamestart Gibson\surnameend}, \bibinfo{author}{Kunal
  \surnamestart Punera\surnameend} \& \bibinfo{author}{Andrew \surnamestart
  Tomkins\surnameend} (\bibinfo{year}{2005}): \emph{\bibinfo{title}{{T}he
  volume and evolution of web page templates}}.
\newblock In \bibinfo{editor}{Allan \surnamestart Ellis\surnameend} \&
  \bibinfo{editor}{Tatsuya \surnamestart Hagino\surnameend}, editors: {\sl
  \bibinfo{booktitle}{{P}roceedings of the 14th {I}nternational {C}onference on
  {W}orld {W}ide {W}eb ({WWW}'05)}}, \bibinfo{publisher}{{ACM}}, pp.
  \bibinfo{pages}{830--839}, \doi{10.1145/1062745.1062763}.

\bibitemdeclare{article}{KadD12}
\bibitem{KadD12}
\bibinfo{author}{Vidya \surnamestart Kadam\surnameend} \&
  \bibinfo{author}{Prakash~R. \surnamestart Devale\surnameend}
  (\bibinfo{year}{2012}): \emph{\bibinfo{title}{A Methodology for Template
  Extraction from Heterogeneous Web Pages}}.
\newblock {\sl \bibinfo{journal}{Indian Journal of Computer Science and
  Engineering ({IJCSE})}} \bibinfo{volume}{3}(\bibinfo{number}{3}).

\bibitemdeclare{inproceedings}{Koh09}
\bibitem{Koh09}
\bibinfo{author}{Christian \surnamestart Kohlsch\"utter\surnameend}
  (\bibinfo{year}{2009}): \emph{\bibinfo{title}{{A} densitometric analysis of
  web template content}}.
\newblock In \bibinfo{editor}{Juan \surnamestart Quemada\surnameend},
  \bibinfo{editor}{Gonzalo \surnamestart Le\'on\surnameend},
  \bibinfo{editor}{Yo\"elle~S. \surnamestart Maarek\surnameend} \&
  \bibinfo{editor}{Wolfgang \surnamestart Nejdl\surnameend}, editors: {\sl
  \bibinfo{booktitle}{{P}roceedings of the 18th {I}nternational {C}onference on
  {W}orld {W}ide {W}eb ({WWW}'09)}}, \bibinfo{publisher}{{ACM}}, pp.
  \bibinfo{pages}{1165--1166}, \doi{10.1145/1526709.1526909}.

\bibitemdeclare{inproceedings}{KohFN10}
\bibitem{KohFN10}
\bibinfo{author}{Christian \surnamestart Kohlsch\"utter\surnameend},
  \bibinfo{author}{Peter \surnamestart Fankhauser\surnameend} \&
  \bibinfo{author}{Wolfgang \surnamestart Nejdl\surnameend}
  (\bibinfo{year}{2010}): \emph{\bibinfo{title}{{B}oilerplate detection using
  shallow text features}}.
\newblock In \bibinfo{editor}{Brian~D. \surnamestart Davison\surnameend},
  \bibinfo{editor}{Torsten \surnamestart Suel\surnameend},
  \bibinfo{editor}{Nick \surnamestart Craswell\surnameend} \&
  \bibinfo{editor}{Bing \surnamestart Liu\surnameend}, editors: {\sl
  \bibinfo{booktitle}{{P}roceedings of the 3th {I}nternational {C}onference on
  {W}eb {S}earch and {W}eb {D}ata {M}ining ({WSDM}'10)}},
  \bibinfo{publisher}{{ACM}}, pp. \bibinfo{pages}{441--450},
  \doi{10.1145/1718487.1718542}.

\bibitemdeclare{inproceedings}{KohN08}
\bibitem{KohN08}
\bibinfo{author}{Christian \surnamestart Kohlsch\"utter\surnameend} \&
  \bibinfo{author}{Wolfgang \surnamestart Nejdl\surnameend}
  (\bibinfo{year}{2008}): \emph{\bibinfo{title}{{A} densitometric approach to
  web page segmentation}}.
\newblock In \bibinfo{editor}{James~G. \surnamestart Shanahan\surnameend},
  \bibinfo{editor}{Sihem \surnamestart Amer-Yahia\surnameend},
  \bibinfo{editor}{Ioana \surnamestart Manolescu\surnameend},
  \bibinfo{editor}{Yi~\surnamestart Zhang\surnameend},
  \bibinfo{editor}{David~A. \surnamestart Evans\surnameend},
  \bibinfo{editor}{Aleksander \surnamestart Kolcz\surnameend},
  \bibinfo{editor}{Key-Sun \surnamestart Choi\surnameend} \&
  \bibinfo{editor}{Abdur \surnamestart Chowdhury\surnameend}, editors: {\sl
  \bibinfo{booktitle}{{P}roceedings of the 17th {ACM} {C}onference on
  {I}nformation and {K}nowledge {M}anagement ({CIKM}'08)}},
  \bibinfo{publisher}{{ACM}}, pp. \bibinfo{pages}{1173--1182},
  \doi{10.1145/1458082.1458237}.

\bibitemdeclare{inproceedings}{NguNPB09}
\bibitem{NguNPB09}
\bibinfo{author}{Dat~Quoc \surnamestart Nguyen\surnameend},
  \bibinfo{author}{Dai~Quoc \surnamestart Nguyen\surnameend},
  \bibinfo{author}{Son~Bao \surnamestart Pham\surnameend} \&
  \bibinfo{author}{The~Duy \surnamestart Bui\surnameend}
  (\bibinfo{year}{2009}): \emph{\bibinfo{title}{A Fast Template-Based Approach
  to Automatically Identify Primary Text Content of a Web Page}}.
\newblock In: {\sl \bibinfo{booktitle}{Proceedings of the 2009 International
  Conference on Knowledge and Systems Engineering}}, {\sl \bibinfo{series}{KSE
  2009}}~, \bibinfo{publisher}{IEEE Computer Society}, pp.
  \bibinfo{pages}{232--236}, \doi{10.1109/KSE.2009.39}.

\bibitemdeclare{inproceedings}{ReiGSL04}
\bibitem{ReiGSL04}
\bibinfo{author}{Davi de~Castro \surnamestart Reis\surnameend},
  \bibinfo{author}{Paulo~Braz \surnamestart Golgher\surnameend},
  \bibinfo{author}{Altigran~Soares \surnamestart Silva\surnameend} \&
  \bibinfo{author}{Alberto Henrique~Frade \surnamestart Laender\surnameend}
  (\bibinfo{year}{2004}): \emph{\bibinfo{title}{{A}utomatic web news extraction
  using tree edit distance}}.
\newblock In: {\sl \bibinfo{booktitle}{{P}roceedings of the 13th
  {I}nternational {C}onference on {W}orld {W}ide {W}eb ({WWW}'04)}},
  \bibinfo{publisher}{{ACM}}, \bibinfo{address}{{N}ew {Y}ork, {NY}, {USA}}, pp.
  \bibinfo{pages}{502--511}, \doi{10.1145/988672.988740}.

\bibitemdeclare{inproceedings}{RowlandsTW09}
\bibitem{RowlandsTW09}
\bibinfo{author}{Tom \surnamestart Rowlands\surnameend}, \bibinfo{author}{Paul
  \surnamestart Thomas\surnameend} \& \bibinfo{author}{Stephen \surnamestart
  Wan\surnameend} (\bibinfo{year}{2009}): \emph{\bibinfo{title}{Web indexing on
  a diet: Template removal with the sandwich algorithm}}.
\newblock In: {\sl \bibinfo{booktitle}{Proceedings of the 14th Australasian
  Document Computing Symposium}}.
\newblock
  \urlprefix\url{http://es.csiro.au/adcs2009/proceedings/poster-presentation/06-rowlands.pdf}.

\bibitemdeclare{inproceedings}{VieSPMCF06}
\bibitem{VieSPMCF06}
\bibinfo{author}{Karane \surnamestart Vieira\surnameend},
  \bibinfo{author}{Altigran~S. \surnamestart da~Silva\surnameend},
  \bibinfo{author}{Nick \surnamestart Pinto\surnameend},
  \bibinfo{author}{Edleno~S. \surnamestart de~Moura\surnameend},
  \bibinfo{author}{Jo\~{a}o M.~B. \surnamestart Cavalcanti\surnameend} \&
  \bibinfo{author}{Juliana \surnamestart Freire\surnameend}
  (\bibinfo{year}{2006}): \emph{\bibinfo{title}{{A} fast and robust method for
  web page template detection and removal}}.
\newblock In: {\sl \bibinfo{booktitle}{{P}roceedings of the 15th {ACM}
  {I}nternational {C}onference on {I}nformation and {K}nowledge {M}anagement
  ({CIKM}'06)}}, \bibinfo{publisher}{{ACM}}, \bibinfo{address}{{N}ew {Y}ork,
  {NY}, {USA}}, pp. \bibinfo{pages}{258--267}, \doi{10.1145/1183614.1183654}.

\bibitemdeclare{inproceedings}{WenHH10}
\bibitem{WenHH10}
\bibinfo{author}{Tim \surnamestart Weninger\surnameend},
  \bibinfo{author}{William \surnamestart Henry~Hsu\surnameend} \&
  \bibinfo{author}{Jiawei \surnamestart Han\surnameend} (\bibinfo{year}{2010}):
  \emph{\bibinfo{title}{{CETR}: {C}ontent {E}xtraction via {T}ag {R}atios}}.
\newblock In \bibinfo{editor}{Michael \surnamestart Rappa\surnameend},
  \bibinfo{editor}{Paul \surnamestart Jones\surnameend},
  \bibinfo{editor}{Juliana \surnamestart Freire\surnameend} \&
  \bibinfo{editor}{Soumen \surnamestart Chakrabarti\surnameend}, editors: {\sl
  \bibinfo{booktitle}{{P}roceedings of the 19th {I}nternational {C}onference on
  {W}orld {W}ide {W}eb ({WWW}'10)}}, \bibinfo{publisher}{{ACM}}, pp.
  \bibinfo{pages}{971--980}, \doi{10.1145/1772690.1772789}.

\bibitemdeclare{inproceedings}{YiLL03}
\bibitem{YiLL03}
\bibinfo{author}{Lan \surnamestart Yi\surnameend}, \bibinfo{author}{Bing
  \surnamestart Liu\surnameend} \& \bibinfo{author}{Xiaoli \surnamestart
  Li\surnameend} (\bibinfo{year}{2003}): \emph{\bibinfo{title}{{E}liminating
  noisy information in Web pages for data mining}}.
\newblock In: {\sl \bibinfo{booktitle}{{P}roceedings of the 9th {ACM} {SIGKDD}
  {I}nternational {C}onference on {K}nowledge {D}iscovery and {D}ata mining
  ({KDD}'03)}}, \bibinfo{publisher}{{ACM}}, \bibinfo{address}{{N}ew {Y}ork,
  {NY}, {USA}}, pp. \bibinfo{pages}{296--305}, \doi{10.1145/956750.956785}.

\end{thebibliography}
\bibliographystyle{eptcs}

\end{document}